\documentclass[conference]{IEEEtran}

\usepackage{url}
\usepackage{hyperref}

\begin{document}

\title{Learning Everywhere: Pervasive Machine Learning for Effective High-Performance Computation}

\author{Geoffrey Fox$^{1}$, James A. Glazier$^{1}$, JCS Kadupitiya$^{1}$, Vikram Jadhao$^{1}$, \\ 
        Minje Kim$^{1}$, Judy Qiu$^{1}$, James P. Sluka$^{1}$, Endre Somogyi$^{1}$, Madhav Marathe$^{2}$, \\ 
        Abhijin Adiga$^{2}$, Jiangzhuo Chen$^{2}$, Oliver Beckstein$^{3}$, Shantenu Jha$^{4}$$^{,5}$\\

\small{\emph{$^{1}$ Indiana University, Bloomington, IN}}\\
\small{\emph{$^{2}$ University of Virginia, Charlottesville, VA}}\\
\small{\emph{$^{3}$ Arizona State University, Tempe, AZ}}\\
\small{\emph{$^{4}$ Rutgers, the State University of New Jersey, Piscataway, NJ 08854, USA}}\\
\small{\emph{$^{5}$ Brookhaven National Laboratory, Upton, New York, 11973}}\\

}


\maketitle
\date{}
\begin{abstract}

The convergence of HPC and data intensive methodologies provide a promising
approach to major performance improvements. This paper provides a general
description of the interaction between traditional HPC and ML approaches and
motivates the "Learning Everywhere" paradigm for HPC. We introduce the concept
of "effective performance" that one can achieve by combining learning
methodologies with simulation based approaches, and distinguish between
traditional performance as measured by benchmark scores. To support the
promise of integrating HPC and learning methods, this paper examines specific
examples and opportunities across a series of domains. It concludes with a
series of open computer science and cyberinfrastructure questions and
challenges that the Learning Everywhere paradigm presents.

\end{abstract}

\section{Introduction}

This paper describes opportunities at the interface between large-scale simulations, experiment design and control, machine learning (ML including deep learning DL) and High-Performance Computing. We describe both the current status and possible research issues in allowing machine learning to  pervasively enhance computational science. How should one do this and where is it valuable? We focus on research challenges on computing for science and engineering (as opposed to commercial) use cases for both big data and big simulation problems. More details  including further citations can be found at \cite{LONGlearningEverywhere}.

The convergence of HPC and data-intensive methodologies \cite{BDHPCConv}
provide a promising approach to major performance improvements. Traditional
HPC simulations are reaching the limits of original progress. The end of
Dennard scaling of transistor power usage and the end of
Moore’s Law as originally formulated has yielded fundamentally different
processor architectures. The architectures continue to evolve, resulting in
highly costly if not damaging churn in scientific codes that need to be finely
tuned to extract the last iota of parallelism and performance.

In domain sciences such as biomolecular sciences, advances in statistical algorithms and runtime systems have enabled extreme scale ensemble based applications \cite{KASSON201887} to overcome limitations of traditional monolithic simulations. However, in spite of several orders of magnitude improvement in efficiency from these adaptive ensemble algorithms, the complexity of phase space and dynamics for modest physical systems, require additional orders of magnitude improvements and performance gains.  

In many application domains, integrating traditional HPC approaches with
machine learning methods arguably holds the greatest promise towards
overcoming these barriers. The need for performance increase underlies the
international efforts behind the exascale supercomputing initiatives and we
believe that integration of ML into large scale computations (for both
simulations and analytics) is a very promising way to get even large
performance gains. Further, it can enable  paradigms such as control or
steering and provide a fundamental approach to coarse-graining which is a
difficult but essential aspect of the many multi-scale application areas.
Papers at two recent workshops BDEC2 \cite{BDEC2process} and NeurIPS \cite{NeurIPS2018} confirm our point
of view and our approach is  synergistic with the BDEC2 process with
its emphasis on new application requirements and their implications for future
scientific computing software platforms.  We would like to distinguish between
traditional performance measured by operations per second or benchmark scores
and the effective performance that one gets by combining learning with
simulation and gives increased performance as seen by the user without
changing the traditional system characteristics. This is of particular
interest in cases where there is a tight coupling between the learning and
simulation components (as outlined below for MLforHPC).
The need for significant enhancement in the effective performance of HPC
motivates the introduction of a new paradigm in HPC: Learning Everywhere!

{\bf Different Interfaces of ML and HPC: } We have identified
\cite{SPIDAL2018A, BDEC2process} several important distinctly
different links between machine learning (ML) and HPC. We define two broad
categories: HPCforML and MLforHPC,
\begin{itemize}
    \item \textbf{HPCforML:} Using HPC to execute and enhance ML performance, or using HPC simulations to train ML algorithms (theory guided machine learning), which are then used to understand experimental data or simulations.
    \item \textbf{MLforHPC:} Using ML to enhance HPC applications and systems
\end{itemize}

This categorization is related to Jeff Dean's "Machine Learning for Systems and Systems for Machine Learning" \cite{Jeff_Dean2017} and Satoshi Matsuoka's convergence of AI and HPC~\cite{Matsuoka2019}.We further subdivide \textbf{HPCforML} as

\begin{itemize}
    \item \textbf{HPCrunsML:} Using HPC to execute ML with high performance
    \item \textbf{SimulationTrainedML:} Using HPC simulations to train ML algorithms, which are then used to understand experimental data or simulations.
\end{itemize}

We also subdivide \textbf{MLforHPC} as

\begin{itemize}
    \item \textbf{MLautotuning:} Using ML to configure (autotune) ML or HPC
simulations. Already, autotuning with systems like ATLAS is hugely successful
and gives an initial view of MLautotuning. As well as choosing block sizes to
improve cache use and vectorization, MLautotuning can also be used for
simulation mesh sizes \cite{NanoIJHPCA} and in big data problems for
configuring databases and complex systems like Hadoop and Spark
\cite{MicrosoftSummit2018A, MicrosoftSummit2018B}.

    \item \textbf{MLafterHPC:} ML analyzing results of HPC as in trajectory
    analysis and structure identification in biomolecular simulations

    \item \textbf{MLaroundHPC:} Using ML to learn from simulations and produce learned surrogates for the simulations. The same ML wrapper can also learn configurations as well as results. This differs from SimulationTrainedML as there typically a learnt network is used to redirect observation whereas in MLaroundHPC we are using the ML to improve the HPC performance.
    \item \textbf{MLControl:} Using simulations (with HPC) in control of experiments and in objective driven computational campaigns \cite{Alexander2018}. Here the simulation surrogates are very valuable to allow real-time predictions. 
\end{itemize}

All 6 topics above are important and pose many research issues in computer
science and cyberinfrastructure, directly in application domains and in the
integration of technology with applications. However, in this paper, we focus
on topics in MLforHPC, with close coupling between ML, simulations, and HPC.
We involve applications as a driver for the requirements and evaluation of the
computer science and infrastructure. In researching {\bf MLaroundHPC} we will
consider ML wrappers for either HPC simulations or complex ML algorithms
implemented with HPC. Our focus is on how to increase effective performance
with the “learning everywhere” principle and how to build efficient “learning
everywhere” parallel systems.



One can view the use of ML learned surrogates as a performance boost that can
lead to huge speedups as calculation of a prediction from a trained network,
can be many orders of magnitude faster than full execution of the simulation
as shown in section \ref{subsec:scaling}.  One can reach Exa or even Zetta
scale equivalent performance for simulations with existing hardware systems.
These high-performance surrogates are valuable in education and control
scenarios by just speeding existing simulations. Simple examples are the use
of a surrogate to represent a chemistry potential or a larger grain size to
solve the diffusion equation underlying cellular and tissue level simulations.
Development of systematic ML-based coarse-graining techniques in both
socio-technical simulations and nano-bio(cell)- tissue layering arises as an
important area of research. In general, Domain-specific expertise will be
needed to understand the needed accuracy and the number of training simulation
runs needed.

There are many groups working in MLaroundHPC but most of the work is just
starting and not built around a systematic study of research issues as we
propose. There is some deep work in building reduced dimension models to use
in control scenarios \cite{Raissi2017A}. We look at three
distinct important areas: Networked systems with socio-technical simulations,
multiscale cell and tissue simulations and at a finer scale biomolecular and
nanoscale molecular systems.

We note that biomolecular and biocomplexity areas which represent 40\% of the
HPC cycles used on NSF computational resources and so this is an area that is
particularly ready and valuable. Molecular sciences has had several successful
examples of using ML for autotuning and ML for analyzing the output of HPC
simulation data. Several fields have made progress in using MLaroundHPC, e.g.,
Cosmoflow and CosmoGAN~\cite{cosmogan} are amongst the
better known projects;  and the Materials community is actively exploring the
uptake of MLControl for the design of materials \cite{BDEC2process}.

This paper does not cover development of new ML algorithms but rather the
advancing the understanding of ML, including Deep Learning (DL) in support of
MLaroundHPC. Of course, the usage experience is likely to suggest new ML
approaches of value outside the MLaroundHPC arena. If one is to use an ML to
replace a simulation, then an accuracy estimate is essential and as discussed
in \ref{sec:UQ} there is a need to build on initial work on UQ (Uncertainty Quantification) with ML \cite{Chan2018} such as
that using dropout regularization to build ensembles for UQ. There are more
sophisticated Bayesian methods to investigate. The research must also address
ergodicity, viz., have we learned across the full phase space of initial
values. Here methods taken from Monte-Carlo arena could be useful as reliable
integration over a domain is related to reliable estimates of values defined
across a domain. Further much of
our learning is for analytic functions whereas much of the existing DL
experience is for discrete-valued classifiers of commercial importance.

Section \ref{sec:CS} discusses cyberinfrastructure and computer science questions, section \ref{sec:UQ} covers uncertainty quantification for learnt results while
section \ref{sec:Infraneeds} the infrastructure requirements needed to implement MLforHPC. Section \ref{subsec:scaling} gives a general performance analysis method and applies to current cases,
Section \ref{sec:Future} covers new opportunities and research issues.

\section{Science Exemplars}

\subsection{Machine learning for Networked Systems}\label{sec:network}

In this section we describe a hybrid method that fuses machine learning and
mechanistic models to overcome the challenges posed by scenarios where data is
sparse and knowledge of underlying mechanism is inadequate. Across domains,
the two approaches have been compared~\cite{peterson2015mechanistic}. Machine
learning approach usually needs a large amount of observation data for
training, and does not explicitly account for mechanisms that govern the the
complex phenomenon. On the other hand, mechanistic models (like agent-based
models) result from a bottom-up approach; but they tend to have too many
parameters, are compute intensive and hard to calibrate. In recent years,
there have been several efforts to study physical processes under the umbrella
of theory-guided data science (TGDS), with focus on artificial neural networks
(ANN) as the primary learning tool. \cite{Karpatne2017} provides a survey of
these methods and their application to hydrology, climate science, turbulence
modeling, etc. where the underlying theory can be used to reduce the variance
in model parameters by introducing constraints or priors in the model space.


Here we consider a particular class of mechanistic
models - network dynamical systems, which have been applied in diverse domains
such as epidemiology and computational social science. A network dynamical
system is composed of a network where nodes of the network are agents
(representing population, computers, etc.) and the edges capture the
interactions between them. A popular example of such systems is the SEIR model
of disease spread in a social network~\cite{newman2002spread}. The complexity
of the dynamics in such a network, due to individual level heterogeneity and
interactions, makes it difficult to train a machine learning model that can be
generalized to patterns not yet presented in historical data. Completely data
driven models cannot discover higher resolution details (e.g. county level
incidence) from lower resolution ground truth data (e.g. state level
incidence).

\noindent {\bf Learning from observational and simulation data: }
Data sparsity is often a challenge for applying machine learning, especially
deep learning methods to forecasting problems in socio-technical systems. One
example of such problems is to predict weekly incidence in future weeks in an
influenza epidemic. In such socio-technical systems, we usually have only
limited observational data, e.g. weekly incidence number reported to the
Centers for Disease Control and Prevention (CDC). Such data is of low spatial
temporal resolution (weekly at state level), not real time (at least one week
delay), incomplete (reported cases are only a small fraction of actual ones),
and noisy (adjusted several times after being published), thus
necessitating a hybrid framework for forecasting by learning from
observational and simulation data.

Observations need to be augmented with existing domain knowledge and behavior
encapsulated in the agent-based model to inform the learning algorithm. In
such hybrid framework, the network dynamical system is used to guide the
learning algorithm so that it conforms to the principles ({\bf consistency}).
At the same time, the learning algorithm will facilitate model selection in a
principled manner. Moreover, the synthetic data goes beyond the observation
data, thus helps voiding overfitting and makes the learned model capable of
processing patterns unseen in the observation data ({\bf generalizability}).
When the dynamical system is more detailed (e.g. individual level) than the
observation data, the hybrid framework allows detailed forecasting ({\bf high
resolution}).

{\bf Epidemic Forecasting: } Simulation trained machine learning methods can
be used for epidemic forecasting. An example of such a  framework is DEFSI
(Deep Learning Based Epidemic Forecasting with Synthetic Information) proposed
in~\cite{Wang2019}. It consists of ($i$) a model configuration module that
estimates a distribution for each parameter in an agent based epidemic model
based on coarse surveillance data; ($ii$) simulation-geenrated synthetic
training data module which generates high-resolution training data by running
HPC simulations parameterized from distributions estimated in the previous
module; ($iii$) a two-branch deep neural network trained on the synthetic
training dataset and used to make details forecasts with coarse surveillance
data as inputs.


Experimental results show that DEFSI performs comparably or better than the
other methods for state level forecasting; and it outperforms the EpiFast
method for county level forecasting. See  Ref. \cite{LONGlearningEverywhere}
and citations therein for details.

\subsection{ML for Virtual Tissue and Cellular Simulations}
\label{sec:VT}
\subsubsection{Virtual Tissue Models}
Virtual Tissue (VT) simulations \cite{Osborne2017} are mechanism-based multiscale spatial simulations of living tissues that  address questions about development, maintenance, damage and repair. They also find application in the design of tissues (tissue engineering) and the development of medical therapies, especially personalized therapies.  VT simulations are computationally challenging for a number of reasons: 1) VT simulations are agent-based, with the core agent often representing biological cells. The number of cells in a real tissue is often of the order of $10^{8}$ or more.  2) Agents are often hierarchical, with agents composed of multiple agents at smaller scales. 3) Agents interact strongly with each other, often over significant ranges \cite{Sluka:2016fz}. 3) Individual agents typically contain complex  sub models that control their properties and behaviors. 4) Materials properties may be complex, like the shear thickening or thinning or swelling or contraction of fiber networks. 5) Modeling transport and diffusion is compute intensive. 6) Models are typically stochastic, so predictivity requires many replicas. 7) Simulations involve uncertainty both in model parameters and in model structure. 8) Biological and medical time-series data are often qualitative, semi-quantitative or differential, making their use in classical optimization difficult. 9) VT models often produce movies of configurations over time. 10) Finally, simulating populations can add several orders of magnitude to the computational challenge. It is possible that ML techniques can be used to short circuit implementations at and between scales.



\subsubsection{Virtual Tissue Modelling and AI + MLandHPC}
AI can directly benefit VT applications in a number of  ways:

\begin{enumerate}
    \item Short-circuiting: The replacement of computationally costly modules with learned analogues.
    \item Parameter fitting in high dimensional parameter spaces. 
    \item Treating stochasticity in results as information rather than noise.
    \item Prediction of bifurcations in models. 
    \item Design of maximally discriminatory experiments --  predict the parameter sets by which two models can be differentiated.
    \item “Run time backwards,” to determine initial conditions that lead to observed endpoints.
    \item The elimination of short time scales, e.g., short-circuit the calculations of advection-diffusion.
    \item Generating additional spatial data sets from experimental images.  
\end{enumerate}



Representative prior work by Karniadakis \cite{Raissi2017A}, Kevrekidis \cite{Kevrekidis2017} and Nemenman \cite{Nemenman2006} shows that neural networks can reproduce the temporal behaviors of biochemical regulatory and signaling networks. Ref. \cite{Liang2017} has shown that networks can learn nonlinear biomechanics simulations of the aorta--being able to predict the stress and strain distribution in the human aorta from the morphology observable with MRI or CT.

\subsection{Machine Learning and Molecular Simulations}
\subsubsection{Nanoscale simulation}
\label{sec:nano}

Despite the employment of the optimal parallelization techniques suited for
the size and complexity of the system, nanoscale simulations remain time
consuming. In research settings, simulations can take up to several days and
it is often desirable to foresee expected overall trends in key quantities;
for example, how does the contact density vary as a function of ion
concentration in nanoscale confinement or how the peak positions of the pair
correlation functions characterizing nanoparticle assembly evolve as the
environmental parameters are tuned. Given the dramatic rise in ML and HPC
technologies, it is not the question of if, but when, ML can be integrated
with HPC to enhance nanoscale simulation meth{}ods. Recent years have seen a
surge in the use of ML to accelerate material simulation techniques: ML has
been used to predict parameters, generate configurations in material
simulations, and classify material properties (see  Ref
\cite{LONGlearningEverywhere} and citations therein). At this time, it is
critical to understand and develop the software frameworks to build ML layers
around HPC to 1) enhance simulation performance 2) enable real-time and
anytime engagement, and 3) broaden the applicability of simulations for both
research and education (in-classroom) usage.

In the context of nanoscale simulation, an initial set of applications for the MLaroundHPC framework can be the prediction of the structure or correlation functions (outputs) characterizing the nanoscale system over a broad range of experimental control parameters (inputs). 
MLaroundHPC can enable the following outcomes:

\begin{enumerate}
    \item Learn pre-identified critical features associated with the simulation output. 
    \item Generate accurate predictions for un-simulated statepoints (by entirely bypassing simulations).
    \item Exhibit auto-tunability (with new simulation runs, the ML layer gets better at making predictions).
    \item Enable real-time, anytime, and anywhere access to simulation results (particularly important for education use).
    \item No run is wasted. Training needs both successful and unsuccessful runs.
\end{enumerate}



To illustrate these outcomes, we discuss nanoscale simulations aimed at the computation of the structure of ions confined by surfaces that are nanometers apart which has been the focus of recent experiments and computational studies 
(see  Ref \cite{LONGlearningEverywhere} and citations therein).
Typically, the entire ionic distribution averaged over sufficient number of independent samples generated during the simulation is a quantity of interest. However, in many important cases, average values of contact density or center density directly relate to important experimentally-measured quantities such as the osmotic pressure \cite{zwanikken1}. Further, often it is useful to visualize expected trends in the behavior of contact or mid-point density as a function of solution conditions or ionic attributes, before running simulations to explore specific system conditions. It is thus desirable that a ``smart'' simulation framework provide rapid estimates of these critical output features with high accuracy. 
MLaroundHPC can enable precisely this as we recently showed that an artificial neural network successfully learns from completed simulation results the desired features associated with the output ionic density profiles to rapidly generate predictions for contact, peak, and center densities in excellent agreement with the results from explicit simulations \cite{NanoICCS}. 

\subsubsection{Biomolecular simulations}
\label{sec:bms}

The use of ML and in particular DL approaches for biomolecular simulations 
\cite{Perez:2018aa} lags behind other areas such as nano-science and materials science \cite{butler2018machine}. This might be partly due to the difficulty to account for large heterogeneous systems with important interactions at short and long length scales. But it might also indicate that the commonly used classical empirical force fields are surprisingly successful \cite{Piana:2014qo} and it is not easy to outperform them at this level of approximation. Therefore, one primary direction of research in this area is to improve the accuracy of the simulation while maintaining the performance of empirical energy functions. 

One promising approach is based on work by Behler and Parrinello 
\cite{behler_generalized_2007} who devised a NN-based potential that was trained on quantum mechanical DFT energies; their key insight was to represent the total energy as a sum of atomic contributions and represent the chemical environment around each atom by an identically structured NN, which takes as input appropriate “symmetry functions” that are rotation and translation invariant as well as invariant to exchange of atoms while correctly reflecting the local environment that determines the energy \cite{behler_first_2017}. Based on this work, Gastegger \textit{et al.} \cite{gastegger_machine_2017} used ML to accelerate ab-initio MD (AIMD) to compute accurate IR spectra for organic molecules including the biological Ala$_{3}^{+}$ tripeptide in the gas phase. Interestingly, the ML model was able to reproduce anharmonicities and incorporate proton transfer reactions between different Ala$_{3}^{+}$ tautomers without having been explicit trained on such a chemical event, highlighting the promise of such an approach to incorporate a wide range of physically relevant effects with the right training data. The ML model was $>$1000 faster than the traditional evaluation of the underlying quantum mechanical physical equations. 

Roitberg \textit{et al.} \cite{s.smith_ani-1:_2017} trained a NN on QM DFT calculations, based on modified Behler-Parrinello symmetry functions. The resulting ANI-1 model was shown to be chemically accurate, transferrable, with a performance similar to a classical force field, thus enabling ab-initio molecular dynamics (AIMD) at a fraction of the cost of "true" DFT AIMD. Extensions of their work with an active learning (AL) approach demonstrated that proteins in an explicit water environment can be simulated with a NN potential at DFT accuracy \cite{smith_less_2018}. The AL approach reduced the amount of required training data to 10\% of the original model \cite{smith_less_2018} by iteratively adding training data calculations for regions of chemical space where the current ML model could not make good predictions. Using transfer learning, the ANI-1 potential was also extended to predict energies at the highest level of quantum chemistry calculations (coupled cluster CCSD(T)), with speedups in the billion.


In general the focus has been on achieving DFT-level accuracy because NN potentials are not cheaper to evaluate than most classical empirical potentials. However, replacing solvent-solvent and  solvent-solute interactions, which typically make up 80\%-90\% of the computational effort in a classical all-atom, explicit solvent simulation, with a NN potential promises large performance gains at a fraction of the cost of traditional implicit solvent models and with an accuracy comparable to the explicit simulations \cite{Wang:2018aa}, as also discussed above in the case of electrolyte solutions. Furthermore, inclusion of polarization, which is expensive (factor 3-10 in current classical polarizable force fields \cite{Lopes:2015aa}) but of great interest when studying the interaction of multivalent ions with biomolecules might be easily achievable with appropriately trained ML potentials.

\section{Integrating ML and HPC: Background and Opportunities}\label{sec:CS}

A primary contribution of this paper is in the categorization, description and
examples of the different ways in which ML can enhance HPC (MLforHPC). Before
we expound upon MLforHPC and open research issues, we provide a a summary
status of HPC for ML (beyond the obvious and well-studied use of GPUs for ML).


\subsection{HPC for Machine Learning}
There has been substantial community progress here with the Industry supported MLPerf \cite{MLPERF} machine learning benchmark activity and Uber’s  Horovod Open Source Distributed Deep Learning Framework for TensorFlow \cite{Horovod}. We have studied  different parallel patterns (kernels) of machine learning applications, looking in particular at Gibbs Sampling, Stochastic Gradient Descent (SGD), Cyclic Coordinate Descent (CCD) and K-means clustering \cite{IPCC-IU}. These algorithms are fundamental for large-scale data analysis and cover several important categories: Markov Chain Monte Carlo (MCMC), Gradient Descent and Expectation and Maximization (EM). We show that parallel iterative algorithms can be categorized into four types of computation models (a) Locking, (b) Rotation, (c) Allreduce, (d) Asynchronous, based on the synchronization patterns and the effectiveness of the model parameter update. A major challenge of scaling is owing to the fact that computation is irregular and the model size can be huge. At the meantime, parallel workers need to synchronize the model continually. By investigating collective vs. asynchronous methods of the model synchronization mechanisms, we discover that optimized collective communication can improve the model update speed, thus allowing the model to converge faster. The performance improvement derives not only from accelerated communication but also from reduced iteration computation time as the model size may change during the model convergence. To foster faster model convergence, we need to design new collective communication abstractions. We identify all 5 classes of data-intensive computation\cite{BDHPCConv}, from pleasingly parallel to machine learning and simulations. To re-design a modular software stack with native kernels to effectively utilize scale-up servers for machine learning and data analytics applications. We are investigating how simulations and Big Data can use common programming environments with a runtime based on a rich set of collectives and libraries for a model-centric approach \cite{Qiu2017C,Qiu2018B}. 

\label{subsec:parallel}

{\bf Parallel Computing: } We know that heterogeneity can lead to difficulty
in parallel computing. This is extreme for MLaroundHPC as the ML learnt result
can be huge factors ($10^5$ in our initial example\cite{NanoICCS}) faster than
simulated answers. Further learning can be dynamic within a job and within
different runs of a given job. One can address by load balancing the unlearnt
and learnt separately but this can lead to geometric issues as quite likely
that ML learning works more efficiently (for more potential simulations) in
particular regions of phase space.

\subsection{Uncertainty Quantification for Deep Learning}
\label{sec:UQ}
An important aspect of the use of a learned ML model is that one must learn not just the result of a simulation but also the uncertainty of the prediction e.g. if the learned result is valid enough to be used. This can be explained in the sense of the bias-variance trade-off, which is based on the decomposition of the expected error into two parts: variance and bias. The variance part explains the uncertainty of the model training process due to the randomness in the training algorithms or the lack of representativeness of the training set. A regularization scheme can reduce the variance so that the model complexity is in control and can result in a {\em smoother} model. However, the regularization approach comes at the cost of an increased amount of bias, which is another term in the expected error decomposition that explains the fitness of the model---by regularizing the model the training algorithm can do only a limited effort to minimize the training error. On the contrary, an unregularized model with a higher model complexity than necessary can also result in a minimal training error, while it suffers from high variance.


Ideally, the bias-variance trade-off can be resolved to some degree by averaging trained instances of an originally complex model. Once these model instances are complex enough to fit the training data set, we can use the averaged predictions as the output of the model. However, averaging many different model instances implies a practical difficulty that one has to conduct multiple optimization tasks to secure a statistically meaningful sample distribution of the predictions. Given the assumption that the model might as well be a complex one to minimize the bias component (e.g. a deep neural network), the model averaging strategy is computationally challenging.

Dropout has been extensively used in deep learning as a regularization technique \cite{dahl2013improving}, but recent researches revisit it as an uncertainty quantification (UQ) tool \cite{gal2016dropout}. The dropout procedure can be seen as an efficient way to maintain a pool of multiple network instances for the same optimization task. It is an efficient ensemble technique as it applies a randomly sampled Bernoulli mask to a layer-wise input unit, thus exposing the optimization process to many differently structured instances of the network. 


A a set of differently thinned versions of the network can form a sample
distribution of predictions to be used as a UQ metric. The dropout-based UQ
scheme can provide an opportunity for the MLaroundHPC simulation experiments.
As a data-driven model it is reasonable to assume that a better ML surrogate
can be found once the training routine sees more examples generated from the
simulation experiment. However, creating more examples to train a better ML
model is a conflicting requirement as the purpose of training the ML surrogate
is to avoid such computation. The UQ scheme can play a role here to provide
the training routine with a way to quantify the uncertainty in the
prediction---once it is low enough, the training routine might less likely
need more data.

\subsection{Machine Learning for HPC}
\label{sec:Infraneeds}
Here we review the nature of the Machine Learning needed for MLforHPC in different application domains. The Machine Learning (ML) load depends on 1) Time interval between its invocations, which will translate into the number of training samples S and 2) size D of data set specifying each sample. This size could be as large as the number of degrees of freedom in simulation or could be (much) smaller if just a few parameters are needed to define simulation. We note two general issues

\begin{itemize} 
\item There can very important data transfer and storage issues in linking the Simulations and Machine Learning parts of system. This could need carefully designed architectures for both hardware and software.
\item The Simulations and Machine Learning subsystems are likely to require different node optimizations as in different types and uses of accelerators.
\end{itemize}

\subsection{Science Exemplar: Nanosimulations}

In this subsection, using the example of Nanosimulations, we show progress
in all areas at the intersection of HPC and ML are having an impact.

In each of two cases below, one is using  scikit-learn, Tensorflow and the Keras wrapper for Tensorflow, as the ML subsystem. The papers \cite{NanoICCS,NanoIJHPCA} are using ML to learn results (ionic density at a given location) of a complete simulation
\begin{itemize} 
    \item D=5 with the five specifying features as confinement length h, positive valency $z_p$, negative valency $z_n$, salt concentration c, and the diameter of the ions d.
\item S=  4805 which 70\% of total 6864 runs with 30\% of the total runs used for testing.
\end{itemize}
In \cite{NanoIJHPCA}, one is not asking ML to predict a result as in \cite{NanoICCS},but  rather training an Artificial Neural Net (ANN) to ensure that the simulation runs at its optimal speed (using for example, the lowest allowable timestep dt and "good" simulation control parameters for high efficiency) while retaining the accuracy of the final result (e.g. density profile of ions). For this particular application, we could get away by dividing a 10 million time-step run (~ 10 nanoseconds that is a typical timescale to reach equilibrium and get data in such systems) into 10 separate runs.

\begin{itemize} 
\item Input data size D= 6 (1 input uses 64 bits floats and 5 inputs use 32 bits integers - total 224 bits)
\item Input number of samples (S) = 15640 (70\% training  30\% test)
\item Hidden layer 1 = 30
\item Hidden layer 1 = 48
\item Output variables = 3
\end{itemize}

Creation of the training dataset took = 64 cores * 80 hrs * 5400 simulation runs = 28160000 or 28 million CPU hours on Indiana University's BigRed2 GPU compute nodes. Each run is 10 million steps long, and you use/learn/train ML every 1 million steps (so block size is a million), yielding 10 times more samples than runs.

Generalizing this, the hardware needs will depend on how often you block, to stop and train the network, and then either on-the-fly or post-simulation, use that training to accelerate simulation or evaluate structure respectively.
Blocking every timestep will not improve the training as typically, it won't produce a statistically independent data point to evaluate any structure you desire. So you want to block at a timescale that is at least greater than the autocorrelation time dc; this is, of course, dependent on example you are looking at -- and so your blocking and learning will depend on the application. In \cite{NanoICCS}, it is small and dc is 3-5 dt; in glasses, it can be huge as the viscosity is high; and in biomolecular simulations, it will also depend on the level of coarse-graining and will be different in fully atomistic or very coarse-grained systems.

The training effort will also depend on the input data size D, and the complexity of the relationship you are trying to learn which change the number of hidden layers and nodes per layer. For example, suppose you are tracking a particle (a side atom on a molecule in a typical nanoscale simulation), in order to come up with a metric (e.g. distance between two side atoms on different molecules) to track the diversity of clusters of particles during the self-assembly process. This comes from expectation that correlations between side atoms may be critical to a macroscopic property (such as formation of these particles into a FCC crystal). In this case your D is huge, and your ML objectives may be looking for a deep relationship, and you may have to invoke an ensemble of ANN's and this will change hardware needs.

\label{subsec:scaling}

{\bf Scaling of Effective Performance: }
An initial approach to estimate speedup in a hybrid MLaroundHPC situation is given in \cite{NanoICCS} for a nano simulation. One can estimate the speedup in terms in terms of four times $T_{seq}$ the sequential execution time of simulation; $T_{train}$ the time for the parallel execution of simulation to give training data; $T_{learn}$ is the time per sample to train the learning networkl; and $T_{lookup}$ is the inference time to predict the results of the simulation by using the trained network. In the formula below, $N_{lookup}$ is the number of trained neural net inferences and $N_{train}$ the number of parallel simulations used in training.
\[Effective Speedup~S= \frac{T_{seq}(N_{lookup}+N_{train})}{T_{lookup}N_{lookup} + (T_{train}+T_{learn})N_{train}}\]
This formula reduces to the classic simple $\frac{T_{seq}}{T_{train}}$ when there is no machine learning and in the limit of large $\frac{N_{lookup}}{N_{train}}$ becomes $\frac{T_{seq}}{T_{lookup}}$ which can be huge! \par
There are many caveats and assumptions here. We are considering a simple case where one runs the $N_{train}$ simulations, followed by the learning and then all the $N_{lookup}$ inferences. Further we assume the training simulations are useful results and not just overhead. We also have not properly considered how to build in the likelihood that training, learning and lookup phases are probably using different hardware configurations with different node counts.

\subsection{Opportunities and Research Issues}\label{sec:Future}





{\bf Research Issues:} In addition to the six categories at the
interface of ML and HPC, the research issues we identify reflect the multiple
interdisciplinary activities linked in our study of MLforHPC, including
application domains described in sections \ref{sec:network}, \ref{sec:VT},
\ref{sec:nano} and \ref{sec:bms}, as well as coarse graining studied in our
case for network science and nano-bio areas.


We have identified the following research areas, which can be categorized into
Algorithms and Methods (1-5), Applied Math (10), Software Systems (6,7),
Performance Measurement and Engineering (8,11).

\begin{enumerate}
    \item Where can application domains use MLaroundHPC and MLautotuning effectively and what science is enabled by this
    \item Which ML and DL approaches are most relevant and how can they be set up to enable broad user-friendly MLaroundHPC and MLautotuning  in domain science
    \item How can Uncertainty Quantification be enabled and separately study ergodicity (bias) and accuracy issues?
    \item Is there new area of algorithmic research focusing on finding algorithms that can be most effectively learnt?
    \item Is there a general multiscale approach using MLaroundHPC.
    \item What are appropriate systems frameworks for MLaroundHPC and MLautotuning. For example, should we wrap microservices invoked by a Function as a Service environment? Where and how should we enable learning systems? Is Dataflow useful?
    \item The different characters of surrogate and “real” executions produce system challenges as surrogate execution is much faster and invokes distinct software and hardware. This heterogeneity gives challenges for parallel computing, workload management  and resource scheduling (heterogeneous and dynamic workflows). The implication for performance is briefly discussed in sections \ref{subsec:parallel} and \ref{subsec:scaling}.

    \item Scaling applications that are composed of multiple heterogeneous computational (execution) units, and have distinct forms of
    parallelism that need balanced performance. Consider a workload comprised
    of $N_L$ learning units, $N_S$ simulations units. The relative number of
    learning units to simulation units will vary with application and problem
    type. The relative values will even vary over execution time of the
    application, as the amount of data generated as a ratio of training data
    will vary. This requires runtime systems that are capable of real-time
    performance tuning and adaptive execution for workloads comprised of
    multiple heterogeneous tasks.

    \item  The application of these ideas to statistical physics problems may need different techniques than those used in deterministic time evolutions.
    \item The existing UQ frameworks based on the dropout technique can provide the level of certainty as a probabilistic distribution in the prediction space. However, it does not always mean that the quality of the distribution is dependent on the quality/quantity of data. For example, two models with different dropout rates can produce different UQ results. If the goal of UQ in MLaroundHPC context is to supply only an adequate amount of data, we need a more reliable UQ method tailored for this purpose rather than the dropout technique that tends to manipulate the architecture of the model.

    \item Application agnostic description and defintion of effective performance enhancement.

\end{enumerate}

\section*{Conclusions}

{\bf Broken Abstractions, New Abstractions:} In traditional HPC the prevailing
orthodoxy is  “Faster is Better” has driven the quest for abstractions of
hierarchical parallelism to speeding up single units of works. Relinquishing
the orthodoxy based upon hierarchical (vertical) parallelism as the only route
to performance is necessary. The new paradigm in HPC --- “Learning
Everywhere”, implies new performance, scaling and execution approaches. In
this new paradigm,  multiple, concurrent heterogeneous units of work replace
single large units of works, which thus require both hierarchical (vertical)
parallelism as well horizontal (many task) parallelism.

\section*{Acknowledgments}
This work was partially supported by NSF CIF21 DIBBS 1443054 and nanoBIO
1720625; the Indiana University Precision Health initiative and Intel through
the Parallel Computing Center at Indiana University. JPS and JAG were
partially supported by NSF 1720625, NIH U01 GM111243 and NIH GM122424. SJ was
partially supported by ExaLearn -- a DOE Exascale Computing project.

\bibliography{draft,nssac,annot,andy,radical,bms}
\bibliographystyle{unsrt}

\end{document}